\documentclass[twocolumn,prl,amsmath,amssymb,showpacs,superscriptaddress,floatfix]{revtex4}
\usepackage{graphicx}
\begin{document}

\title{Charge transport through a SET with a mechanically oscillating island}

\author{N.~M. Chtchelkatchev}
\affiliation{Departement Physik und Astronomie, Universit\"at
Basel, Klingelbergstr. 82, 4056 Basel, Switzerland}
\affiliation{L.D.\ Landau Institute for Theoretical Physics,
Russian Academy of Sciences, 117940 Moscow, Russia}
\affiliation{Institute for High Pressure Physics, Russian Academy
of Sciences, Troitsk 142092, Moscow Region, Russia}

\author{W. Belzig}
\affiliation{Departement Physik und Astronomie, Universit\"at
Basel, Klingelbergstr. 82, 4056 Basel, Switzerland}

\author{C. Bruder}
\affiliation{Departement Physik und Astronomie, Universit\"at
Basel, Klingelbergstr. 82, 4056 Basel, Switzerland}

\date{\today}

\begin{abstract}
We consider a single-electron transistor (SET) whose central island is
a nanomechanical oscillator. The gate capacitance of the SET depends
on the mechanical displacement, thus, the vibrations of the island 
vibrations may strongly influence the current-voltage
characteristics, current noise, and higher cumulants of the current.  
Harmonic oscillations of the island and oscillations with random
amplitude (e.g., due to the thermal activation) change the transport
characteristics in a different way. The noise spectrum has a peak at
the frequency of the island oscillations; when the island oscillates
harmonically, the peak reduces to a $\delta$-peak. We show that
knowledge of the SET transport properties helps to determine in
what way the island oscillates, to estimate the amplitude, and the
frequency of the oscillations.
\end{abstract}


\maketitle

The interplay of electric currents through nanostructures with their
mechanical degrees of freedom has attracted a lot of interest
recently, both from the experimental and theoretical side
\cite{cleland,roukes,gorelik,blick,zwerger,cond-mat/0312239,Blanter,armour1,mitra}.
One of the central questions of this new
field of nanophysics is how
the vibrations of the oscillating part of a nanodevice influence
its transport properties and vice versa. A number of
nanomechanical devices were investigated in the last years, e.g.,
so-called single-electron shuttles \cite{gorelik,blick,zwerger}. 
On the theoretical side, it was shown recently that electric
currents passing through a dirty nanowire can stimulate its
vibrations \cite{shytov}. 
Indications for thermal vibrations of suspended single-wall
nanotubes doubly clamped between two contacts were observed \cite{babic}.

Recently, the nanomechanical properties of single-electron transistors
(SETs) have started to attract attention. SETs built from carbon
nanotubes are a natural candidate for this type of question. For
instance, it was shown that the equilibrium shape of a suspended
nanotube studied as a function of a gate voltage shows features
related to single-electron electronics, e.g., Coulomb ``quantization''
of the nanotube displacement \cite{Blanter}.

In this Letter, we discuss how vibrations of the central island of the
SET change the current and the noise. We show that the
transport characteristics of the SET differ for islands oscillating
thermally or harmonically. The current noise spectrum has a peak at the
frequency of the island oscillations that reduces to a $\delta$-peak
when the island oscillates harmonically. Therefore, knowing the
transport properties of the SET can help to determine in what way the
island oscillates, and to find the amplitude and frequency of the
oscillations.

\begin{figure}[thb]
\includegraphics[width=60mm]{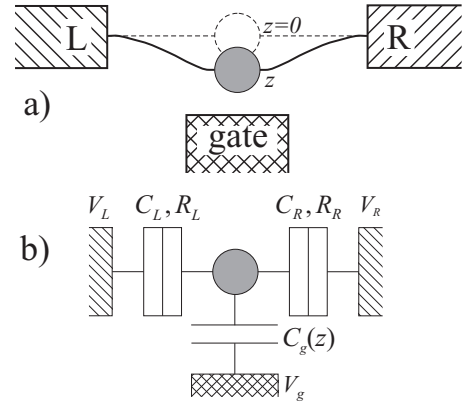}
\caption{\label{fig1} (a) Sketch of a single-electron transistor
(SET) with an oscillating island. The island is coupled to the left
(L) and right(R) leads by tunnel junctions, and its capacitance to the
gate $C_g(z)$ depends on the coordinate $z$ that measures the
deviation of the island from its equilibrium position.
(b) Equivalent circuit of the device.}
\end{figure}

The system that we want to study -- a SET with a mechanically
oscillating island -- is sketched in Fig.~\ref{fig1}. We assume
that the island is coupled to the left (L) and right (R) leads by
tunnel junctions but can mechanically vibrate.
A suspended nanotube is a possible experimental realization of this
scenario \cite{Blanter}.

The charge of the island is coupled to the leads and the gate through
the capacitances $C_L$, $C_R$, and $C_g(z)$; $z$ is the
transverse deviation of the island from its equilibrium position.
When the island oscillates, the gate capacitance changes with the position
of the island center $z$, and therefore the transport properties of the SET
change.

We assume that electronic transport through the SET can be described
by sequential tunneling. In this case, it is governed by four tunneling
rates \cite{beenakker91,averin91,Hershfield}: the rate for 
electrons to tunnel onto the
central region from the left ($\Gamma^L_{n\to n+1}$) and right
($\Gamma^R_{n\to n+1}$) and the rates for electrons to tunnel off
the central region. The rates can be
calculated via Fermi's golden rule. The energy change
corresponding to the first tunnel process is
\begin{gather}\label{Delta_E}
\Delta E^L_{n\to n+1}=e(V_g-V_L)+\varepsilon_n,
\\ \notag
\varepsilon_n=\frac{e^2}{C_\Sigma}\left[\frac 1
2+n+\frac{C_R(V_R-V_g)+C_L(V_L-V_g)}e\right]\; ,
\end{gather}
where $C_\Sigma=C_L+C_R+C_g(z)$.  Defining
$\gamma(\varepsilon)\equiv-\epsilon/(1-\exp(\beta\epsilon))$, the
rates can be written as 
\begin{gather}\label{Gamma}
\Gamma^L_{n\to n+1}(z)=\frac 1 {e^2R_L}\gamma(\Delta E^L_{n\to
n+1}(z))\; ,
\end{gather}
where $R_L$ is the resistance of the left junction. The other rates can
be written similarly. The charge state of the SET is characterized by
the probability $\rho_n(t)$ to find $n$ excess electrons on the island.
The time evolution of $\rho_n(t)$ is governed by the master equation
\cite{beenakker91,averin91,Hershfield,BruderSchoeller94}:
\begin{gather}\label{Master_Equation}
\frac {\partial{\rho}_i} {\partial t}=\sum_{a=\pm
1}(\Gamma_{i+a\to i}\rho_{i+a}-\Gamma_{i\to i+a}\rho_i)\; ,
\end{gather}
where $\Gamma_{i\to j}=\Gamma_{i\to j}^R+\Gamma_{i\to j}^L$.
The current and all its cumulants can be
expressed through $\Gamma$ and $\rho$ \cite{Bagrets}.

Equation~(\ref{Master_Equation})has to be supplemented with the equation
describing the oscillations of the island. The electrostatic force
that acts on the island \cite{Landau8} is  
$f(z)=\partial_z C_g(V_g-\varphi)^2/2$, where
$\varphi=(C_LV_L+C_RV_R+C_GV_G+q)/C_\Sigma$ is the potential of
the island, and $q=\sum_n\rho_n ne$ the average charge. Then the motion of
the island can be described by a Langevin equation,
\begin{gather}\label{Diff}
\ddot{z}+\eta \dot{z}+\omega_0^2z=y+[f(z)-f(0)]/m\; ,
\end{gather}
where $m$ is the island mass;
$\eta\sim \omega_0/Q$ where $Q$ is the quality factor; finally, $y$ is a
random force simulating the interaction with the thermal bath,
$\langle y(t) y(t')\rangle=2\eta m T\delta(t-t')$
\cite{Landau5}. Typically $Q\sim 10^3-10^4$ 
\cite{cleland}
and $\eta\ll \omega_0$.
If the island is a nanotube, $C_g$ is a functional of the deviation
$z(x)$ where $x$ is the coordinate along the nanotube \cite{Blanter}.
The rates depend only on integral quantities like $\int z(x)
dx/L$ ($L$ is the length of the nanotube). Their dynamics can be
described by Eq.~(\ref{Diff}) if the amplitude of the nanotube
oscillations does not exceed its diameter by several orders of magnitude.

The current going through the left junction \cite{Hershfield} is
\begin{gather}
I(t)=e\sum_{j} [\Gamma_{j\to j+1}^L(t)-\Gamma_{j\to j-1}^L(t)]\rho_j(t)\; .
\end{gather}
We are interested in the current averaged over a time interval
$\tau$ much larger than the characteristic period $T_0$ of the
island oscillations: $\bar{I}=\int_{-\tau}^\tau
I(t,z(t))dt/2\tau$, $\tau\to \infty$.

The typical frequency of micromechanical oscillations is
$\omega_0\sim 100$MHz. If electrons tunnel through the SET with a
similar frequency, the current will be of the order of $I\sim
e\omega_0\sim 10^{-11}$A. However, the current going through a typical
SET is usually several orders of magnitude bigger. Hence, during the
oscillation period $T_0$ many electrons go through the SET and thus it
is possible to neglect the time derivative in the master equation
\eqref{Master_Equation}.  Then all the methods used for current
calculation in ``usual'' SETs are applicable to the case with the
oscillating island \cite{Hershfield}. Averaging the current over time
can be replaced by averaging over $P(z)$, the density of the
probability distribution for the deviation $z$, i.e., $\bar{I}=\int
P(z) I(z)dz$. If the island oscillations are thermally activated,
$P(z)\propto \exp(-z^2/2\langle z^2\rangle_T)$, where $\langle
z^2\rangle_T=k_BT/m\omega_0^2$. If the island oscillates harmonically,
$z(t)=z_0\sin(\omega t)$, then $P(z)=1/\pi\sqrt{1-(z/z_0)^2}$. In
these expressions, the driving terms $\sim f$ in Eq.~(\ref{Diff}) that
couple the current in the SET with its mechanical degrees of freedom
were neglected. This term is usually much smaller than $\omega_0^2 z$ 
on the left-hand side of Eq.~(\ref{Diff})
(e.g., for the SET parameters in recent experiments, see
Ref.~\onlinecite{Blanter}): the small parameter
is $z_0\max_i{\partial_z\ln C_i}$.  The driving terms
may become important, e.g., when an a.c. bias near the resonance
frequency $\omega_0$ is applied to the terminal(s) of the SET. Then the
a.c. bias drives the island oscillations and the $f$-terms in Eq.~(\ref{Diff})
are important to determine the amplitude $z_0$ of the oscillations
\cite{Usmani-Blanter_unpublished}.

\begin{figure}[t]
\includegraphics[width=80mm]{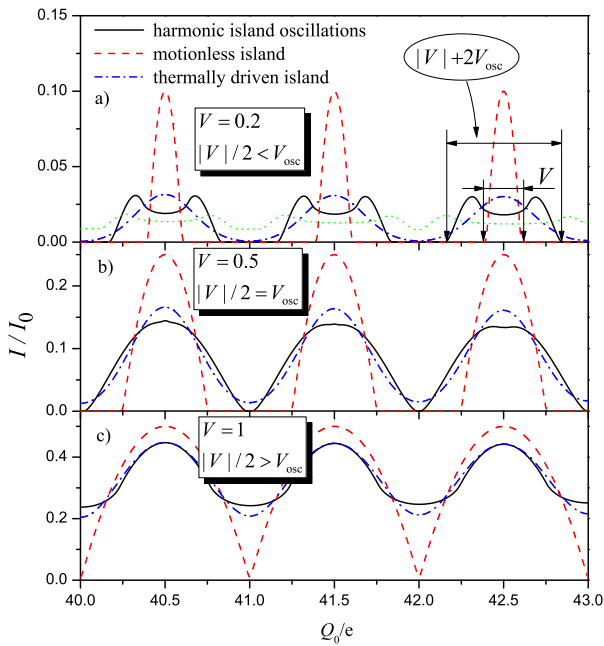}
\caption{\label{fig2} Current gate-charge characteristics for
  symmetric SET ($R_L=R_R$) with $C_g\gg C_{L,R}$, $V_L=-V_R=V/2$,
  $Q_0=-C_g^{(0)}V_g$.  (a) $V=0.2$, ($|V|/2<V_{\rm osc}$), (b) $V=0.5$,
  ($|V|/2\approx V_{\rm osc}$), (c) $V=1$, ($|V|/2> V_{\rm osc}$). The
  voltage is measured in units of $|e|/2C_\Sigma(z=0)$, the current in
  units of $I_0=(e/C_g^{(0)})/(R_L+R_R)$. The dashed curve corresponds
  to a static island, the solid curve to a harmonically
  oscillating island, $z=z_0\sin(\omega_0 t)$;
  $z_0(\partial_z C_g)/C_g =5.6\cdot 10^{-3}$ (this
  is typical for SETs where the island is a nanotube \cite{Blanter};
  then $z_0\approx 5r$, where $r$ is a typical nanotube diameter). 
  The dotted curve in Fig.~\ref{fig1}a illustrates what happens if
  $V_{\rm osc}/(|e|/2C_g^{(0)})=5>1$ and Eq.~(\ref{I_z_expansion}) is
  not valid. The dash-dotted curves correspond to the case of thermal
  motion; the thermal average $\langle z^2\rangle_T\equiv
  k_B T/m\omega_0^2$ is chosen to be equal $\langle (z_0\sin(\omega_0
  t))^2\rangle_t=z_0^2/2$. The integer part of $Q_0/e$ is the number of
  electrons on the island in the static regime when $V_L=V_R=0$. The
  curves are periodic in the static case, but not if the island
  oscillates. The areas under the peaks in the static and in the
  dynamical cases are the same.}
\end{figure}
In general, the current and noise in a SET cannot be calculated
analytically for arbitrary transport voltages \cite{Ingold-Nazarov}
even when the island is static. 
Analytical progress can be made if  
we restrict ourself to the case of
small driving voltages near the onset voltage, 
and temperatures much below the charging energy $e^2/C_{\Sigma}$,
i.e., $\gamma(\varepsilon)\approx\theta(\varepsilon)$ in
Eq.~\eqref{Gamma}.  In the case of
a static island, the performance of the SET as a transistor and
electrometer reaches an optimum in this regime \cite{Ingold-Nazarov}. In
this region, the transport characteristics of the SET are also most
sensitive to mechanical oscillations of the island, so this regime is
the most interesting. Only two states of the island have to be taken
into account; the probability $\rho$ has only two nonzero values
$\rho_n,\rho_{n+1}$ \cite{Ingold-Nazarov}. If $V_L<V_R$ an electron
enters the island with the rate $\Gamma_{n\to n+1}^L(z)$ from the left
lead and goes away with the rate $\Gamma_{n+1\to n}^R(z)$ into the right
lead. The average current will be
\begin{gather}\label{I_classical}
    \bar{I}=\int  dz P(z)
    \frac{e\Gamma_{n\to n+1}^L(z)\Gamma_{n+1\to n}^R(z)}
    {\Gamma_{n\to n+1}^L(z)+\Gamma_{n+1\to n}^R(z)}\; ,
\end{gather}
where $n=[-C_g^{(0)}V_g/e]$, $C_g^{(0)}\equiv C_g(z=0)$, and 
$[\ldots]$ means the integer part.
Assuming that the capacitances depend only weakly on $z$, 
Eq.~(\ref{I_classical}) can be expanded with respect to $z$. To
proceed,  we define
\begin{gather}
    J(z)=\frac{1}{e} \frac{\Delta E^L_{n\to n+1}\Delta E^R_{n+1\to
    n}}{R_R\Delta E^L_{n\to n+1}+R_L\Delta E^R_{n+1\to
    n}}\; .
\end{gather}
Using
$\Delta E^R_{n+1\to n}=e(V_R-V_L)-\Delta E^L_{n\to n+1}$ and 
defining $z_1,z_2$ to be the roots of the
equations $\Delta E^L_{n\to n+1}=0$ and $\Delta E^R_{n+1\to n}=0$, 
we get from Eq.~(\ref{I_classical})
\begin{multline}\label{I_z_expansion}
\bar{I}\approx\int_{\min(z_1,z_2)}^{\max(z_1,z_2)}  dz P(z)[J(0)+
\\
z(\partial_z J(z)|_{z\to 0})+z^2\frac 1 2(\partial_z^2 J(z)|_{z\to 0})]\; .
\end{multline}
This formula is valid also for $V_L>V_R$.
Using Eq.~\eqref{Delta_E} we find 
$z_2-z_1 \approx e(V_R-V_L)/\partial_z\varepsilon_n|_{z\to 0}$.
Thus, if $z_0$ is a characteristic amplitude of island oscillations then
it is natural to define the voltage scale
\begin{gather}\label{Vosc_general}
    V_{\mathrm {osc}}=\frac{z_0}{e}(\partial_z\varepsilon_n|_{z\to
0})\; .
\end{gather}
In the limiting case $C_L,C_R\ll C_g^{(0)}$,
\begin{gather}\label{Vosc}
V_{\mathrm{osc}}=\frac{z_0\partial_z C_g|_{z\to 0}}{[C_g^{(0)}]^2}e(n+1/2)\; .
\end{gather}
Eq.~\eqref{I_z_expansion} is valid if
$V_{\mathrm{osc}}<e/C_{\Sigma}$. If the driving voltages applied
to the SET terminals are much larger than $V_{\rm osc}$, the
integration limits in Eq.~\eqref{I_z_expansion} can be extended to infinity
because they far exceed $z_0$, the scale of decay of $P(z)$. 
The second term in Eq.~\eqref{I_z_expansion} vanishes and
\begin{gather}\label{I_correction1}
\bar I=I(z=0)+ \frac 1 2 \langle z^2\rangle\frac{\partial^2
}{\partial z^2}I(z)|_{z\to 0}\; ,
\end{gather}
where $\langle z^2\rangle=\int P(z) z^2dz$. The first term in
Eq.~\eqref{I_correction1} is the current for a static island. If
the driving voltages applied to SET terminals are smaller than
$V_{\rm osc}$ then the second term (linear in $z$) in
Eq.~\eqref{I_z_expansion} does not vanish; in this regime the
current-voltage characteristics is largely influenced by island
oscillations. The small parameter in the expansion 
Eq.~\eqref{I_z_expansion} is $z_0\partial_z C_g$. 
The second term in Eq.~\eqref{I_correction1} is of
second order in this parameter. If we investigate the $I-V$
characteristics, island oscillations will smear the Coulomb gap
within a voltage band of width of order $V_{\rm osc}$.

\begin{figure}[t]
\includegraphics[width=80mm]{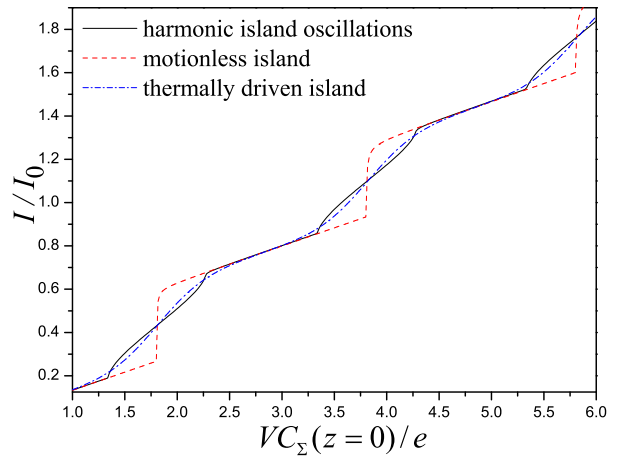}
\caption{\label{fig_staircase} Coulomb staircase, $R_R/R_L=1000$,
  $V_L=-V_R=V/2$, $C_L=C_R=C_g^{(0)}$, $Q_0=40.6$ ($C_g^{(0)}$ and other
  parameters are as in Fig.~\ref{fig2}).  The average square amplitude
  of the island deviations is equal to that of the harmonically
  oscillating island. The Coulomb steps are smeared on the
  scale $V_{\rm osc}$. For other choices of $V_g$ the staircase graph
  looks similarly.}
\end{figure}
The $I-V_g$ characteristics of a symmetric SET 
($R_L=R_R$, $C_L=C_R\ll C_g$) with an oscillating island is shown in
Fig.~\ref{fig2}.  The dashed curves correspond to the case of
the static island. The solid and dash-dotted curves show the case of a
harmonically oscillating island or
an island subject to thermal equilibrium fluctuations. 
Figures~\ref{fig2}(a-c) also show
how the $I-V_g$ characteristics change when the driving voltage $V$ is
smaller, of the order of, or larger than $V_{\rm osc}$. Within each
peak of the curves, $n$ is constant, therefore $V_{\rm osc}$ is
also constant within the peak. The most interesting case is shown in the first
panel of Fig.~\ref{fig2}(a).
For the harmonically oscillating island, the peaks split and their
width becomes larger with the characteristic scale $V_{\rm osc}$; when
the island moves due to thermal activation the peaks do not split
but their width also becomes larger with $V_{\rm osc}$. Thus, the type
of motion of the island leaves a characteristic trace in the 
$I-V_g$ plot. Equation~\eqref{I_z_expansion} well describes $I-V_g$
characteristics when the peaks do not overlap, like in
Figs.~\ref{fig2}a and b. It follows from Eq.~\eqref{I_classical} that
the areas under the peaks the in static and dynamical cases are
equal. The $I-V_g$ characteristics is periodic in the static case, but
not periodic for an oscillating island because $V_{\rm osc}$
changes from peak to peak, see, e.g., Eq.~\eqref{Vosc}.
For this graph, we used the parameters of the nanotube model of
the island (see Ref.~\onlinecite{Blanter}); we chose, e.g,
$C_g(z)=L/2\ln(2(R-z)/r)$, where $r=0.65$nm and  $L=500$nm are the
nanotube radius and length, and $R=100$nm is the distance to the
gate.  If $C_L, C_R$ and $C_g$ are of the same order, the $I-V_g$
characteristics is qualitatively similar to what is shown in
Fig.~\ref{fig2}.
\begin{figure}[t]
\begin{center}
\includegraphics[width=90mm]{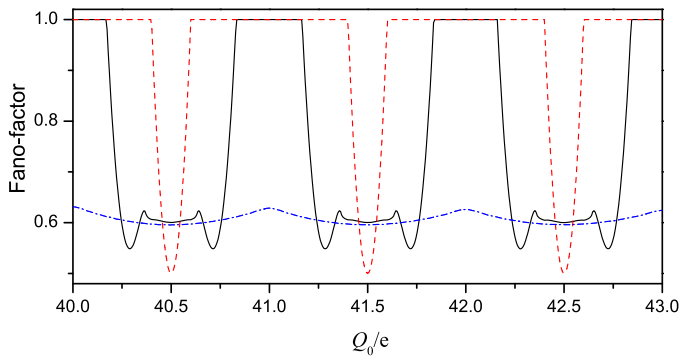}\end{center}
\caption{\label{fig_noise} Fano factor for $V<V_{\mathrm{osc}}$. 
The dashed line corresponds to
the static case, the solid line to the harmonically
oscillating island, and the dash-dotted line to the thermally driven island.
The average square amplitude of the island oscillations are equal
for the harmonically oscillating island and the case of thermal equilibrium.}
\end{figure}
If the junction is asymmetric, $R_R\neq R_L$, the peaks in 
Fig.~\ref{fig2} become nonsymmetric. Figure~\ref{fig_staircase} 
shows the $I-V$ characteristics for $R_R\gg R_L$. The oscillations of
the island smear the Coulomb gap within a voltage range of the order of
$V_{\mathrm{osc}}$.

We will now discuss the current noise in a SET with a moving island.
When the island oscillates, the irreducible current-current correlator
$S(\tau,\Theta)=
\langle\langle I_L(\Theta+\tau/2)I_L(\Theta-\tau/2)\rangle\rangle$
depends on both $\tau$ and $\Theta$ (rather than only on $\tau$ as in the
case of the static island). However, since the charging events in the SET
are correlated on time scales much shorter than the period of the
island oscillations, the dependence of $S$ on $\tau$ is much stronger
than on $\Theta$, and the zero-frequency noise can be found as
$\overline{\int S(\tau,\Theta)d\tau}$, where the bar means averaging over
$\Theta$. In other words, the low-frequency noise can be calculated at a
given position of the island (e.g., as it is done in
Ref.~\onlinecite{Hershfield}) and then averaging over time as it was
already done for the current above. The result of this procedure is
presented in Fig.~\ref{fig_noise} which shows the dependence of the
Fano factor \cite{Blanter-Buttiker} on $V_g$. Here we assumed that the
driving voltage is smaller than $V_{\rm osc}$, i.e., the system is in
the region in which the influence of the oscillations of the island on
the transport properties of the SET is maximal.  The Fano factor dips
get split due to harmonic oscillations of the island; the scale of the
splitting is $V_{\rm osc}$. In contrast to that, the dips are washed
out by thermal equilibrium oscillations of the island.
Here, we assumed that the noise frequency is much below the frequency of the
island oscillations, $\omega_0$. 

We now consider the noise spectral density, i.e., the Fourier transform
$S(\omega,\Theta)$ of $S(\tau,\Theta)$.  
If $\omega$ approaches $\omega_0$, the
correction to the noise from the motion of the island is
$\sim(\partial_z I)^2\langle\langle z^2\rangle\rangle_\omega$, where
$\langle\langle z^2\rangle\rangle_\omega$ is the irreducible correlator
of the deviation $z$ at frequency $\omega$. It has a $\delta$-peak at
$\omega_0$ if the island oscillates harmonically.  In contrast, if the
island moves due to thermal activation, $\langle\langle
z^2\rangle\rangle_{\omega}=2\eta k_BT/
m((\omega^2-\omega_0^2)^2+\omega^2\eta^2)$ \cite{Landau5}, the
noise peak has a width of the order of the oscillation 
damping factor $\eta$, see Eq.~(\ref{Diff}).  A similar result was
obtained numerically in Ref.~\onlinecite{armour2}. Thus, measuring the noise
spectrum allows to find the frequency of island oscillations and gives
information on the nature of the oscillations.

In conclusion, we have discussed how vibrations of the island in a SET
change its transport properties.  The transport characteristics of the
SET can be used to determine the nature of island motion,
in particular, to estimate the amplitude and frequency of its
oscillations.

We would like to thank Ya.~M. Blanter and C. Sch\"onenberger for
stimulating discussions.  This work was financially supported by RFBR
and the Russian Ministry of Sciences (N.M.C.), 
by the Swiss NSF, and the NCCR Nanoscience.

\end{document}